\documentclass[12pt]{JHEP3}
\usepackage{amsmath,epsfig}
\usepackage{amsfonts,amssymb,amsthm}
\def\be{\begin{equation}}
\def\ee{\end{equation}}
\def\ba{\begin{eqnarray}}
\def\ea{\end{eqnarray}}

\newtheorem{lemma}{Lemma}
\newtheorem{proposition}[lemma]{Proposition}
\newtheorem{theorem}[lemma]{Theorem}

\def\be{\begin{equation}}
\def\ee{\end{equation}}
\def\bea{\begin{eqnarray}}
\def\eea{\end{eqnarray}}

\def\yzero{\smash{\hbox{$y\kern-4pt\raise1pt\hbox{${}^\circ$}$}}}

\def\be{\begin{equation}}
\def\ee{\end{equation}}
\def\bea{\begin{eqnarray}}
\def\eea{\end{eqnarray}}

\def\-{\hphantom{-}}

\def\s2{\frac{1}{\sqrt2}}

\def\beq{\begin{equation}}
\def\eeq{\end{equation}}
\def\beqa{\begin{eqnarray}}
\def\eeqa{\end{eqnarray}}

\def\Tr{{\rm Tr \,}}

\def\IF{\relax{\rm I\kern-.18em F}}
\def\II{\relax{\rm I\kern-.18em I}}
\def\IP{\relax{\rm I\kern-.18em P}}
\def\IC{\relax\hbox{\kern.25em$\inbar\kern-.3em{\rm C}$}}
\def\IR{\relax{\rm I\kern-.18em R}}

\def\Dsl{\,\raise.15ex\hbox{/}\mkern-13.5mu D} 
\def\IZ{Z\kern-.4em  Z}

\def\R{{\cal R}}

\title{On the Spectral properties of Multi-branes, M2 and M5 branes}

\author{M. P. Garc\'\i a del Moral $^1$, A. Restuccia $^3$\,\footnote{E-mail:
\emph{garciamormaria@uniovi.es;arestu@usb.ve}}\\
$^1$ Departamento de F\'\i sica, Universidad de Oviedo, Avda Calvo
Sotelo S/n. Oviedo, Espa\~na\\
$^3$ Departamento de F\'\i sica, Universidad de Antofagasta, Aptdo 02800, Chile \\
 $\&$ Departamento de F\'\i sica, Universidad Sim\'on Bol\'\i var\\
Apartado 89000, Caracas 1080-A, Venezuela}

\abstract{In this note we summarize some of the properties found in \cite{amilcar},\cite{gmmnpr}\cite{bgmr4}. We characterize spectral properties of the quantum mechanical hamiltonian of theories with fermionic degrees of freedom beyond semiclassical approximation.  We obtain a general class of bosonic polynomial potentials for which the Schr\"oedinger operator has a discrete spectrum. This class includes all the scalar potentials in membrane, 5-brane, p-branes, multiple M2 branes, BLG and ABJM theories. We also give a sufficient condition for discreteness of the spectrum for supersymmmetric and non supersymmetric theories with a  fermionic contribution. We characterize then the spectral properties of different theories: the BMN matrix model,  the supermembrane with central charges and a bound state of $N$ D2 with $m$  D0. We show that, while the first two models have a purely discrete spectrum with finite multiplicity, the latter has a continuous spectrum starting from a constant given in terms of the monopole charge.}

\preprint{FPAUO-11/05}

\begin{document}
\section{Introduction}
An important aspect of super-membranes, super 5-branes and
supersymmetric multiple-M2 branes refers to the quantum stability of the theory and the validity of the Feynman
kernel. A natural way to proceed is to formulate the theory on a compact base manifold, perform then a
regularization of the theory in terms of an orthonormal basis and
analyze properties of the spectrum of the associated Schr\"odinger
operator. This procedure, to start with a field theory and analyze its properties
by going to a regularized model, has been very useful in
 field theory, although relevant symmetries of the theory may be lost
  in the process. The quantum properties for a large class of regularized model, in particular the ones we are discussing, is then determined from the Schr\"odinger
    operator $H=-\triangle+V(x)+\textrm{fermionic terms}$ with $V(x)=\sum_i\Big[P_i(x)\Big]^2,\label{I}$ where
$P_i(x)$ is a homogeneous polynomial on the configuration variables
$x\in\R^n$. For example, in the membrane theory $P_i(x)$ are of degree two.
A knowledge of the complete spectrum encodes information about the higher order interacting terms beyond the semiclassical approximation. The first few bound states provide information about the potential in neighborhoods of the point on the configuration space at which potential is minimum, while the nature of the spectrum is related to the behavior of the potential at large distances in the configuration space. In turns, the latter is closely connected to properties of the physical model at high energies. Discreteness of the spectrum with accumulation point at infinity for the Hamiltonian of a physical theory
renders a compact resolvent. Mathematically this is an amenable property
as far as the study of the high energy eigenvalues is concerned. On the
one hand, this guarantees the existence of a complete set of
eigenfunctions, which can be used to decompose the action of the operator
in low/high frequency expansions. On the other hand, the study of
eigenvalue asymptotics for the resolvent (or the corresponding heat
kernel) in the vicinity of the origin, can be carried out by means of the
so-called Schatten-von Neumann ideals. None of this extends in
general, if the Hamiltonian has a non-empty essential spectrum. See
\cite{br}. Physical theories like YM or SYM among others, when formulated in a box of diameter $L$ have discrete spectrum \cite{luscher} while in the $L$ going to infinite limit, they exhibit a continuous spectrum with, in some cases, a mass gap. For supermembranes, 5-branes and multiple-superbranes the situation is different, when formulated on a compact base manifold, the spectrum of their regularized hamiltonian is continuous, generically from $[0,\infty]$. The analysis of interacting Hamiltonian for such theories becomes complicated since there is no guarantee that their Feynman formulation provides the correct kernel of the theory. A good example of this situation is the BFSS theory \cite{bfss}. Besides, even when there is a well defined semiclassical approximation for the theory, its properties cannot be extrapolated to the interacting theories. For example a discrete spectrum at the semiclassical level does not imply, in general, a discrete spectrum of the interacting Hamiltonian.
\section{Useful Criteria to characterize Bosonic Spectra}
Is there a precise condition on the potential which characterizes the discreteness of the spectrum for a  bosonic matrix model?
This was achieved
by A.~M.~Molchanov \cite{Molchanov:1953am} and more recently
extended by V.~Maz'ya and M.~Schubin \cite{Maz'ya:2005v}. It makes
use of the mean value of the potential, in the sense of Molchanov, on
a star shaped cell $\mathcal{G}_d$, of diameter $d$. The spectrum
is discrete if and only if the mean value of the potential goes to
infinity when the distance from $\mathcal{G}_d$ to a fixed point on
configuration space goes to infinity in all possible ways. The
potential is assumed to be locally integrable and bounded from
below. Using the above theorem we proved in \cite{amilcar} the following proposition which allows to show that all bosonic membrane, multibrane and p-brane Hamiltonia have discrete spectrum. The proof may also be obtained from the results from \cite{rs4} which are very useful for polynomial potentials. We intend to use these results also for non polynomial perturbations of membranes and multi-brane theories, for that reason the Molchanov approach seems more appropriate.
\begin{proposition}\label{a1}
Let $H=-\Delta +V(X)$ be a Schr\"{o}edinger operator with potential $V(X)$
given by%
\small{\begin{equation}
V(X)=\sum\limits_{M_{1},...,M_{l}}\sum\limits_{B=1}^{N}\left(
X_{M_{1}}^{a_{1}}...X_{M_{l}}^{a_{l}}f_{a_{1}...a_{l}}^{B}\right)
^{2} \label{E14}
\end{equation}}%
let $\mathcal{M}$ be the symmetric matrix defined in (\ref{E12}),
$\left[ X_{M_{i}}^{a_{i}}\right] \in \mathbb{R}^{M\times N}$ and
$f_{a_{1},...,a_{l}}^{B}$ real coefficients satisfying the following
restriction: $\mathcal{M}$ is strictly positive definite. Then $H$
is essentially self adjoint and has a discrete spectrum in
$L^{2}\left( \mathbb{R}^{M\times N}\right) $.
\end{proposition} The components of the matrix $\mathcal{M}$ are given by:
\begin{equation}
\mathcal{M}_{a\widehat{a}}:=\mathcal{F}_{a;\widehat{a}%
}=f_{c_{1},...,c_{l-1},a}^{B}f_{c_{1},...,c_{l-1},\widehat{a}%
}^{B}+....+f_{c_{1},...,c_{i-1},a,c_{i+1},...c_{l-1}}^{B}f_{c_{1},...,c_{i-1},%
\widehat{a},c_{i+1},...,c_{l-1}}^{B}+...  \label{E12}
\end{equation} It is by construction positive and the requirement on the proposition is that it is strictly positive.

\subsection{ BLG case}
To characterize  the non-perturbative spectral properties of
the scalar bosonic potential of BLG/ABJM type , it is necessary first, to
formulate these theories in the regularized matrix formalism. These
theories have real scalar fields  $X^{aI}$ valued in the
bifundamental representation of the $\mathcal{G}\times \mathcal{G'}$
algebra, gauge fields $A_{\mu}^{ab}$ where
    $\mu=0,1,2$ spanning the target-space dimensions, and $a\in
    \mathcal{G},b\in \mathcal{G^{'}}$,
     and spinors $\Psi_{a\alpha }$ also valued in the algebra.
     Let us consider the sixth degree scalar potential of the BLG case,
\small{ \begin{equation} V=\int dx^3
    \frac{1}{12}Tr([X^I,X^J,X^K])^2\end{equation}}
where $f_d^{efg}$ are the 'structure constants' of the algebra color
generators $T_a$. For the BLG
case a 3-algebra relation is satisfied
$[T^a,T^b,T^c]=f^{abc}_{d}T^d.$
We expand now each of the fields $X^{I}_a$ in a basis of generators
$T_A$. To obtain the regularized model,
$ X^{I}_a= \sum X^{IA}_aT_A$ with
     $A=(a_1,a_2)$. For the enveloping algebra of $su(N$),
 $T_AT_B=h_{AB}^CT_C$,$\eta_{AB}=\frac{1}{N^4}Tr{(T_AT_B)}$ where $h_{AB}^C$
    are given in \cite{dwhn},\cite{gmr}.
     To obtain the regularized model, the potential can be re-written as a squared-term \begin{equation}
V=\frac{1}{12}(F^{\mathcal{A}\mathcal{B}\mathcal{C}}_{\mathcal{U}}X^{\mathcal{A}I}X^{\mathcal{B}J}X^{\mathcal{C}K})^2\end{equation}
with coefficients
$F^{\mathcal{A}\mathcal{B}\mathcal{C}}_{\mathcal{U}}=f^{abc}_{u}
h_{AB}^{E}h_{CE}^{U}$ that do not exhibit antisymmetry in the
indices $\mathcal{A}=(A,a),\mathcal{B}=(B,b),\mathcal{C}=(C,c)$ nor
are structure constants. Using the Proposition \ref{a1} we can
assure that this regularized potential has a purely discrete
spectrum since
$
f^{abcd}X_a^{IA}=0 \to X_a^{IA}=0.\Box
$
 We thus have that the D=11 membrane, the 5-brane, p-branes as well as the bosonic BLG model satisfy the ssumptions of the Proposition \ref{a1} and hence their regularized versions have discrete spectrum \cite{gmmnpr}\cite{mnpr}.

\subsection{ ABJM case}

ABJM theory \cite{abjm} can be obtained from the 3-algebra expression by
relaxing some antisymmetric properties of the 3-algebra structure
constant as it is indicated in \cite{Bagger:2008se} considering now
instead of real scalar fields, -as happens in the BLG case-, complex
ones $Z^{a\alpha }$ . In the ABJM case
\cite{abjm}, the bosonic scalar potential may be re-expressed in a
covariant way as a sum of squares \cite{Bandres:2008ry}. Using the
results of \cite{Bagger:2008se} where the potential is
\small{\[V = \frac{2}{3}\Upsilon^{CD}_{Bd}\bar\Upsilon_{CD}^{Bd}\]}
where $\Upsilon^{CD}_{Bd} =
f^{ab\overline{c}}{}_dZ^C_aZ^D_b\overline{Z}_{B\overline{c}}
-\frac{1}{2}\delta^C_Bf^{ab\overline{c}}{}_dZ^E_aZ^D_b\overline{Z}_{E\overline{c}}+\frac{1}{2}\delta^D_B
f^{ab\overline{c}}{}_dZ^E_aZ^C_b\overline{Z}_{E\overline{c}}.$
The zero-energy solutions correspond to $\Upsilon^{CD}_{Bd}=0$. In
distinction with the case of BLG, the ABJM potential includes a sum
of three squared terms. The indices $C,D$ are mandatory different
but not necessarily the index $B$. We can bound the potential for
the one with
$
\Upsilon^{CD}_{B^{'}d} = f^{ab\overline{c}}{}_dZ^C_a Z^D_b
\overline{Z}_{B^{'}\overline{c}}$ where $B^{'}$ is an index different from $C,D$. To reduce the analysis to one in quantum mechanics,  a regularization
procedure is performed.  The regularity condition of Proposition
\ref{a1}, in terms of the triple product \cite{Bagger:2008se}, may
be expressed as a \begin{equation}\label{a70}
[X,T^b;\overline{T}^{\overline{c}}]=f^{ab\overline{c}}{}_d X_a T^d
=0 \quad \forall{b,\overline{c}}\Rightarrow X_a=0.\end{equation}

 Note that if this condition is not satisfied, the
potentials we are considering have continuous spectrum. This result
follows using Molchanov, Maz'ya and Schubin theorem. Factorizing out the constants due to regularization process, in the case of
ABJM and ABJ it follows from (49) in \cite{Bagger:2008se} that
(\ref{a70}) implies
$
(t^{\lambda}_{\alpha})^{a \overline{c}}X_a=0,
$ where $t^{\lambda}_{\alpha}$ are $u(N)$ representations of the gauge algebra $\mathcal{G}$.
 In the case $\mathcal{G}$ is $u(N)$ then the regularity condition is satisfied.
The proposition (\ref{a1}) in our paper ensures then that the
Schr\"oedinger operator associated to the regularized scalar sixth
degree potential of ABJM has also purely discrete spectrum.$\Box$

\section{ Hamiltonians with a Fermionic Contribution}
The analysis of these hamiltonians now gets much more complicated. In fact the Molchanov, Mazya-Shubin theorem cannot be applied directly because supersymmetric potentials are not bounded from below. Suppose that in $L^2(\mathbb{R}^{N})\otimes \mathbb{C}^{d}$, the
operator realization of the Hamiltonian has the form $
   H= P^2 + V(Q)$, $Q\in \mathbb{R}^N$
where $V$ is a Hermitean  $d\times d$ matrix whose entries are continuous
functions of the configuration variables $Q$.
 Assume additionally that $V(Q)$ is bounded from below by $b$, that is $
    V(Q)w\cdot w \geq b |w|^2$ $w\in \mathbb{C}^d$
where $b\in \mathcal{R}$ is a constant. Then $H$ is bounded from below by $b$ and the spectrum of $H$ does not intersect the interval $(-\infty,b)$.
The following abstract criterion established conditions guaranteeing that the spectrum of $H$ is purely discrete.
The proof of these lemmas can be found in \cite{bgmr}and \cite{bgmr4}.

\begin{lemma} \label{criterion} Let $v_k(Q)$ be the
eigenvalues of the $d\times d$ matrix $V(Q)$.
If all $v_{k}(Q)\to +\infty$ as $|Q|\to \infty$, then the spectrum
of $H$ consists of a set of isolated
eigenvalues of finite multiplicity accumulating
at $+\infty$.
\end{lemma}

\begin{lemma} \label{nuevo_lemma}
Let $V_B$ be a continuous bosonic potential of the configuration space.
Let $V_F$ be a fermionic matrix potential with continuous entries $v_{ij}$
of the configuration space.
Suppose that there exist constants
$b_B,b_F,R_0,p_B,p_F>0$ independent of $Q\in\mathcal{R^N}$
satisfying the following conditions
\[
   V_B \geq b_B |Q|^{p_B} \qquad \text{and} \qquad  |v_{ij}|\leq b_F |Q|^{p_F}
\]
for all $|Q|>R_0$. If $p_B>p_F$, then the Hamiltonian
$H=P^2+ V(Q)$ of the quantum system associated with $V=V_BI+V_F$
has spectrum consisting exclusively
of isolated eigenvalues of finite multiplicity.\end{lemma}


\subsection{The BMN matrix model} \label{section3}
The  matrix model for the Discrete Light Cone Quantization (DLCQ)
of M-theory on the maximally supersymmetric pp-waves background of eleven dimensional supergravity examined in \cite{bmn}
fits in well with the abstract framework of lemmas 2.5 and 2.6. The dynamics of this theory is described by
 the following $U(N)$ matrix model, which in our notation reads $\small{L_{BMN}=T-V_B-V_F}$
\small{\begin{equation}
V_B=\Tr \left[
\frac{\mu^2}{36R}\sum_{i=1,2,3}(X^i)^2+\frac{\mu^2}{144R}\sum_{i=4}^9
(X^i)^2+\frac{i\mu}{3}\sum_{i,j,k=1}^3\epsilon_{ijk}X^iX^jX^k-\frac{R}{2}\sum_{i,j=1}^9[X^i,X^j]^2\right]\\
\end{equation}}
The quartic contribution to the potential
 with an overall minus sign is positive, since the
  commutator is antihermitean. The coordinates
    $X^i$, for $i=4,\dots,9$, only contribute quadratically and quartically to the Lagrangian,
therefore, they satisfy the bound of Lemma~\ref{nuevo_lemma}, with $p_B=2$ and $p_F=1$. Thus, the analysis of the
    bosonic potential may be focus in the first three coordinates.
The potential vanishes at the variety determined by the condition
$
[X^i,X^j]=\frac{i\mu}{6R}\epsilon^{ijk}X^k.
$
with the rest of the fields equal to zero.
In turns, this condition corresponds to a fuzzy sphere, \cite{bmn}, along the
directions $1,\,2$ and $3$, so there are no flat directions with zero
potential. Let us now analyze the potential away from the minimal set
in the configuration space. To characterize completely the system let  $\rho^2=\sum_{i=1}^3
Tr(X^{i})^2$  and $\varphi\equiv \frac{X}{\rho}$ be defined on a unitary
hypersphere $S^{3N^2}$. Let $V_{B1}=\frac{\mu^2}{36R}\rho^2 P(\rho,\varphi)$

\begin{theorem} \label{lemma_D0D2}
Let $R_0>\frac{\mu}{3R}\sqrt{C_2(N)N}$ where $C_2(N)=\frac{N^2-1}{4}$ and $\mu,R$ different from zero.
Then $P(\rho,\varphi)>C>0$ for all $\rho>R_0$ and $\varphi\in S^{3N^2}$.
\end{theorem}To analyze the supersymmetric contribution, we just have to realize that the fermionic contribution
 is linear in the bosonic variables, so it satisfies the assumptions of
 Lemma~\ref{nuevo_lemma}. Consequently the
 supersymmetric spectrum of the BMN matrix model has the following remarkable property also shared with the
 supermembrane with central charges. Its Hamiltonian has a purely discrete spectrum with eigenvalues of finite multiplicity
only accumulating at infinity. We emphasize that the spectrum is discrete  in the whole real line. It should be noted however that, at present, there are not clear restrictions about the spectrum of the model in the large $N$ limit, in fact $R_{0}\to\infty$ when $N\to\infty$. In principle, it might have a complicated continuous spectrum with the presence of gaps.
\subsection{ The supermembrane with central charges}\label{section4}
 The action of the supermembrane with central charges \cite{bgmr2}, with
base manifold a compact Riemann surface $\Sigma$ and Target Space $\Omega$
the product of a compact manifold and a Minkowski space-time, is
defined in term of maps: $\Sigma \longrightarrow \Omega$,
satisfying a certain topological restriction over $\Sigma$. This
restriction ensures that the corresponding maps are wrapped in a
canonical (irreducible) manner around the compact sector of
$\Omega$. In order to generate a nontrivial family of admissible maps,
this sector is not arbitrary but rather it is constrained by the
existence of a holomorphic immersion $\Sigma\longrightarrow
\Omega$.
In particular, let $\Sigma$ be a torus and
$\Omega=T^2\times M_9$ where $T^2=S^1\times S^1$ is the flat
torus. Let $X_r:\Sigma \longrightarrow T^2$ with $r=1,2$ and
$X^m:\Sigma\longrightarrow M_9$ with $m=3,\ldots,9.$ The
topological restriction is explicitly given in this case by the
condition
$
\epsilon_{rs}\int_{\Sigma} dX_r\wedge dX_s= n \operatorname{Area}(\Sigma)\ne
0.$
the theory is formulated in 9D and its hamiltonian is the following:
\begin{equation}\label{e}
\begin{aligned}
H_{d}=&\int \sqrt{w}d\sigma^{1}\wedge d\sigma^{2}[\frac{1}{2}(\frac{P_{m}}{\sqrt{W}})^{2}
+\frac{1}{2}(\frac{\Pi^{r}}{\sqrt{W}})^{2}+\frac{1}{4}\{X^{m},X^{n}\}^{2}+\frac{1}{2}(\mathcal{D}_{r}X^{m})^{2}
+\frac{1}{4}(\mathcal{F}_{rs})^{2}\\ \nonumber & +
 \Lambda(\{\frac{P_{m}}{\sqrt{W}},X^{m}\}-\mathcal{D}_{r}\Pi^{r}
]+\int_{\Sigma} \sqrt{W}[-\overline{\Psi}\Gamma_{-}\Gamma_{r}\mathcal{D}_{r}\Psi
+ \overline \Gamma_{-}\Gamma_{m}\{X^{m},\Psi\}]+\Lambda \{\overline{\Psi}\Gamma_{-}, \Psi\}.
\end{aligned}
\end{equation}
where  \small{$\mathcal{D}_r X^{m}=D_{r}X^{m} +\{A_{r},X^{m}\}$},
\small{$\mathcal{F}_{rs}=D_{r}A_s-D_{s }A_r+ \{A_r,A_s\}$},
 $D_{r}=2\pi
R^{r}\frac{\epsilon^{ab}}{\sqrt{W}}\partial_{a}\widehat{X}^{r}\partial_{b}$
and $P_{m}$ and $\Pi_{r}$ are the conjugate momenta to $X^{m}$ and
$A_{r}$ respectively. $\Psi$ are Majorana spinors.
$\mathcal{D}_{r}$ and $\mathcal{F}_{rs}$ are the covariant
derivative and curvature of a symplectic noncommutative theory
\cite{ovalle}, \cite{mr} constructed from the symplectic
structure $\frac{\epsilon^{ab}}{\sqrt{W}}$ introduced by the central
charge. The integral of the curvature we take it to be constant and
the volume term corresponds to the value of the hamiltonian at its
ground state. The physical degrees of the theory are the $X^{m},
A_{r},\Pi^{c}, \Psi$. They are single valued fields on $\Sigma$.
We consider two different regularization schemes, one is a bottom-up approach , the $SU(N)$ regularization, other one, more appropriate for the large N analysis is a top-down regularization and is a cut-off in the number of degrees of freedom irrespective of the symmetries lost in the regularize model. In both cases we obtain by means of Lemmas \ref{criterion}, ~\ref{nuevo_lemma}, that the spectrum is purely discrete as it is proved in the original papers \cite{bgmr,bgmr4}.
The first of these regularization of the supermembrane with central charges was proposed in \cite{gmr}. It is invariant under infinitesimal transformations generated by the first class constraint obtained by variations on $\Lambda$ of the Hamiltonian below. This first class constraint satisfies an $SU(N)$ algebra.
In \cite{bgmr4}we verified that the potential satisfies the bound of
the Lemma~\ref{nuevo_lemma}. The argument follows in analogous fashion as in the previous case although the proof is much more complicated and we refer to the interested reader to the original papers. We may write the potential as
$V=\frac{n^2}{16\pi^2 N^3}\rho^2 P(\rho,\varphi)$, in terms of a hypersphere of radius $\rho$  and angles $\varphi$ with
$\rho$=$\sum_{m,r}\Tr\left( \frac{1}{N}[T_{V_r},X^m]T_{-V_r}\right)^2+\sum_{s,r}\left(\frac{1}{N}[T_{V_s},A_r]T_{-V_s}\right)^2$ and $
\varphi=\left(\frac{X^m}{\rho},\frac{A_r}{\rho}\right)$ and perform a proof based on the Lemmas \ref{criterion},\ref{nuevo_lemma}. From this it follows that the spectrum is purely discrete.
\subsection{ D2-D0 system: Nonempty essential spectrum}
We now consider a model which describes the reduction of a 10D $U(N)$ Super Yang-Mills to (1+0) dimensions, however we allow the presence of monopoles. Consider a (2+1) Hamiltonian whose bosonic contribution is given by
\small{\[
H=\int_{\Sigma}\Tr\left[ \frac{1}{2}((P^m)^2+(\Pi^i)^2)+\frac{1}{4}\left(F_{ij}^2+2(D_i X^m)^2+(i[X^m,X^n])^2\right)\right]
\]}
which satisfies the monopole condition $
\int_{\Sigma}\Tr F=2\pi m,\quad m\in\mathbb{N}.
$
We decompose the $U(N)$ valued 1-form  $\hat{A}$ as $\hat{A}=aD+A$ with $A\in SU(N)$,
 and $D\in U(N)$. The monopole condition is then $\int_{\Sigma} d a=2\pi\frac{m}{\Tr D}$.We then write the $1+0$ Hamiltonian $H=\frac{1}{2}\tilde{H}$,
$
\tilde{H}=-\Delta +V_B+V_F,
$
where
\small{
\begin{align*}
V_B
=\frac{1}{2}\Tr \left[ (i[X^m,X^n])^2+2(i[X^m,\hat{A}_i])^2+F_{ij}^2 \right], \quad
 F_{ij}= \frac{mD\epsilon_{ij}}{\Tr D}+ i[\hat{A}_i,\hat{A}_j]
 \end{align*}}
 and $V_F$ is the supersymmetric Yang-Mills fermionic potential.
In order to analyze the spectrum, we observe that there are directions escaping to infinity at which $V_B$ remains finite. In fact, in any direction at which all the brackets vanish, the wave function can escape to infinity with finite energy. This means that the spectrum has necessarily a continuous sector. In order to construct precisely a wave function in the corresponding $L^2$ space, $\Psi=\Psi_F\phi_0\chi$,
 we introduce $X=\frac{1}{d+2}(\sum_m X^m +\sum_i A_i)$
 where the range of $m$ is $d$ and  the range of $i$ is $2$ and define $\widetilde{X}^m=X^m-X$ and $\widetilde{A}_i=\hat{A}_i-X.$ where $M=1,\dots,d+2$, $\widetilde{X}^{d+1}=\widetilde{A}_1$ and $\widetilde{X}^{d+2}=\widetilde{A}_2.$
We simplify the argument by taking $D$ to be diagonal. We may then use the gauge freedom of the model to impose that $X$ is also diagonal.
Following \cite{dwln}, the above allows us to construct a sequence of wave functions
which happen to be a singular Weyl sequence for any $E\in [\frac{1}{2}\frac{m^2 \Tr D^2}{(\Tr D)^2},+\infty)$.
These ``pseudo-eigenfunctions'' are the product of  compactly supported cutoff
  $\chi_t\equiv \chi(\left\|x\right\|-t,\frac{X}{\left\|X\right\|},\widetilde{X}^M_{||})$ normalized by the condition
$\int_{\widetilde{X}_{||}^m, X} \chi^2_t=1$. It is a wave function with support moving off to infinity as $t\to\infty$, a fermionic wavefunction $\Psi_F$ and the bosonic $L^2$ function
 \small{\small{ \[
\phi_0=\left(\left\|X\right\|^l\frac{
\det{g_{ab}}}{\pi^l}\right)^\frac{1}{4}
 \exp\left(-\frac{\left\|x\right\|}{2}(\widetilde{X}^{Ma}g_{ab}\widetilde{X}^{Mb})\right),
\quad\int_{\widetilde{X}_{T}^m}\phi_0^2=1\]}}
where $l=(d+2)(N^2-N)$. Note that $X^{M}=\widetilde{X}^{Ma}T_a$ where $T_a$ are the generators of $U(N)$ and $g_{ab}$ is the square root of the positive symmetric matrix $(d+2)(f_{ab}^c\frac{X^b}{\left\|X \right\|}f_{cd}^e\frac{X^e}{\left\|X \right\|}).$
The normalized fermionic wave function is the limit when $t\to\infty$ of the eigenfunction of the fermionic interacting term, associated to the negative eigenvalue with highest absolute value. One can now evaluate $\lim_{t\to\infty}(\Psi,H\Psi)$ . In this limit the only term of $V_B$ that does not vanish is the constant one. There is a cancelation of the quadratic terms in $X$ between the contribution of the Laplacian and that of  the potential, also the linear term in $\left\|X\right\|$, arising from the action of the Laplacian on $\phi_0$, is exactly cancelled by the fermionic eigenvalue which is also linear in $\left\|X\right\|$. This is a supersymmetric effect. Although the monopole in this case breaks supersymmetry, the cancelation occurs exactly as in the model without monopoles. The resulting consequence of this is that
\small{\[
\lim_{t\to\infty}(\Psi,H\Psi)=\int_{\widetilde{X}_{||},X} \chi_t(-\Delta_x-\Delta_{\widetilde{X_{||}}})\chi_t +\frac{1}{2}m^2\frac{\Tr D^2}{(\Tr D)^2}.
\]}
 One may choose $\chi$ such that the first term is equal to any scalar $E\in[0,\infty)$. The spectrum of the original Hamiltonian is therefore continuous and it comprises the interval $[\frac{1}{2}\frac{m^2 \Tr D^2}{(\Tr D)^2},+\infty).$

\section{Conclusion}
We discuss on the results in \cite{gmmnpr}, where a general proof for the discreteness of scalar bosonic, polynomial matrix models including: M2, M5, p-branes, BLG, ABJ/M was obtained. In fact, a continuous spectrum at the regularized bosonic model arising from a formulation on a compact space, would imply several difficulties on the models. For example, the Feynmann kernel would be ill defined.  This is the first step in order to consider a non perturbative analysis of these new models. We also characterize the spectrum of three models containing fermionic sectors. We analyze a D2-D0 (N,K) system. Irrespectively to the numbers of N  D2's, K D0's it has continuous spectrum starting from a valued determined by the monopole contribution. For the supersymmetric multibrane models (BLG, ABJ/M) we estimate that the  spectrum is continuous and has a mass gap. Then we analyze beyond semiclassical approximation two models of the supermembrane corresponding to different backgrounds on the target-space that in distinction with the 11D regularized supermembrane \cite{dwhn},\cite{dwln} have discrete spectrum at regularized level: the BMN matrix model which arises as a DLCQ of the supermembrane on a pp-wave\cite{dsjvr} and the Supermembrane with a topological condition. For the BMN matrix model we conjecture a non-empty essential  spectrum at the continuum limit. The supermembrane with central charges in two different regularizations has purely discrete spectrum. The large N limit of our bound converge to the value we already found in \cite{bgmr2} in the large N limit. Moreover, the regularized  semiclassical eigenvalues converge properly to the semiclasical ones in the continuous. The bosonic potential of the full  theory also satisfy a analogous type of bound that the regularized one \cite{bgmr2}. For all these evidences it seems plausible that the large N limit of this theory will have a purely discrete spectrum with finite multiplicity with accumulation point at infinity. Iff  that is the case, the supermembrane with central charges could be interpreted as a fundamental supermembrane describing microscopical degrees of freedom of (at least part) of M-theory.
\section{Acknowledgements}
A. Restuccia thanks the organizers of the Workshop `XVI European Workshop pn String Theory 2010' celebrated in Madrid from 14-18 of June, Spain for the possibility to present this work.  The work of MPGM is funded by
the Spanish Ministerio de Ciencia e Innovaci\'on (FPA2006-09199) and
the Consolider-Ingenio 2010 Programme CPAN (CSD2007-00042).

\end{document}